\documentclass{article}

     \PassOptionsToPackage{numbers, compress}{natbib}

\usepackage[final]{neurips_2022_ml4ps}

\usepackage[utf8]{inputenc} 
\usepackage[T1]{fontenc}    
\usepackage{hyperref}       
\usepackage{url}            
\usepackage{booktabs}       
\usepackage{amsfonts}       
\usepackage{nicefrac}       
\usepackage{microtype}      
\usepackage{xcolor}
\usepackage{graphicx}
\usepackage{bbm}
\usepackage{amsmath}
\usepackage{booktabs}
\usepackage{todonotes}
\usepackage{siunitx}

\title{Emulating Fast Processes in Climate Models}

\author{
Noah D. Brenowitz\thanks{Now at NVIDIA Research: \url{https://research.nvidia.com/person/noah-brenowitz}} \\
  Allen Institute for Artificial Intelligence \\
  \texttt{nbren12@gmail.com} \\ 
  \And
  W. Andre Perkins \\
  Allen Institute for Artificial Intelligence \\
  \texttt{andrep@allenai.org} \\
 \And
 Jacqueline M. Nugent \\
 University of Washington \\
 \texttt{jnug@atmos.washington.edu}
 \And
       Oliver Watt-Meyer \\
  Allen Institute for Artificial Intelligence \\
  \texttt{oliverwm@allenai.org} \\ 
  \And
  Spencer K. Clark \\
  Allen Institute for Artificial Intelligence,\\Geophysical Fluid Dynamics Laboratory, NOAA \\
  \texttt{spencerc@allenai.org} \\
  \And
Anna Kwa \\
  Allen Institute for Artificial Intelligence \\
  \texttt{annak@allenai.org} \\
      \And
  Brian Henn \\
  Allen Institute for Artificial Intelligence \\
  \texttt{brianhenn@allenai.org} \\
   \And
  Jeremy McGibbon \\
  Allen Institute for Artificial Intelligence \\
  \texttt{jeremym@allenai.org} \\ 
    \And
  Christopher S. Bretherton \\
  Allen Institute for Artificial Intelligence \\
  \texttt{christopherb@allenai.org} \\ 
}

\begin{document}

\maketitle
\begin{abstract}
Cloud microphysical parameterizations in atmospheric models describe the formation and evolution of clouds and precipitation, a central weather and climate process. Cloud-associated latent heating is a primary driver of large and small-scale circulations throughout the global atmosphere, and clouds have important interactions with atmospheric radiation.
Clouds are ubiquitous, diverse, and can change rapidly. In this work, we build the first emulator of an entire cloud microphysical parameterization, including fast phase changes. The emulator performs well in offline and online (i.e. when coupled to the rest of the atmospheric model) tests, but shows some developing biases in Antarctica. Sensitivity tests demonstrate that these successes require careful modeling of the mixed discrete-continuous output as well as the input-output structure of the underlying code and physical process.
\end{abstract}

\section{Introduction}

Weather forecasts and climate simulations  are predominantly made by evolving atmospheric and oceanic fluid dynamics forward in time on a discrete global grid.
Most atmospheric models are organized into a dynamical core, which evolves the Navier-Stokes equations on the rotating sphere, and the ``physics'' which handle everything else such as cloud formation, rain, and boundary layer turbulence.

Replacing climate model physical parameterizations with machine learning models is known as emulation.
Emulation provides a principled path towards accelerating Fortran model codes, especially for use on computing architectures reliant on accelerators such as GPUs which can efficiently run many machine learning (ML) applications. 
Because of this, most emulation studies have focused on radiative transfer \citep{Chevallier1998-su,Krasnopolsky2005-ca}, which is the slowest sub-component in the typical atmospheric physics suite.
However, recent studies have emulated deep convection \citep{OGorman2018-hn}, gravity wave drag \citep{Chantry2021-az}, atmospheric chemistry \citep{Keller2019-qn,Kelp2022-lb,Schreck2022-na}, and the warm rain process \citep{Gettelman2021-ej}. 
In spite of these successes, no general purpose approach to emulation has emerged and consistently high accuracy over all timescales and spatial locations has been elusive.

Emulation is also an excellent test-bed for more complex ML approaches to improve physical parameterizations since it is very clearly posed as a supervised learning task.
It can be argued that successfully emulating a given climate model component is a necessary first step before trying to improve that component e.g by training models with fine resolution data \citep{Brenowitz2019-qs,Rasp2018-ff,Yuval2020-ks,Bretherton2022-ex}.
This work often focuses on moist atmospheric physics, a notoriously difficult process to parameterize.
One key moist process is the cloud microphysics scheme, which handles fast phase changes like condensation, evaporation and precipitation.
Microphysics are tightly coupled to the dynamical core through the corresponding latent heat release.
Since cloud microphysics are computationally cheaper than radiation, no study has yet tried to emulate an entire microphysics scheme including phase changes. However, in view of its importance and its contrasting character to radiation, this is a worthy and educational ML challenge.

In this study, we train ML models to emulate an simple microphysical scheme written in Fortran, including the phase changes.
We will show that our best scheme has excellent offline skill and works well, but not perfectly, online when coupled to the rest of the atmospheric solver.

\section{Methods}\label{sec:methods}

The training data are generated by running the FV3GFS  weather model \citep{Zhou2019-xh,Harris2021} with the simplified Zhao-Carr (ZC) microphysics (Appendix \ref{sec:zc}), for simplicity.
The ZC microphysics predict changes in cloud liquid and ice condensate, precipitation, and associated heating and moistening rates at each grid point in a grid column given a corresponding column of thermodynamic inputs.
The change of temperature $T$, cloud $c$, or humidity $q$ by the ZC microphysics is defined as $\Delta = \Delta_g + \Delta_p$.
$\Delta_g$ is the change to due the grid-scale condensation subroutine (\verb+gscond+).
$\Delta_p$ is the change due the precipitation subroutine (\verb+precpd+).
We added hooks before and after these subroutines that are used both for saving the data and applying the ML emulators online.

FV3GFS  is a compressible atmospheric model used for operational weather forecasts by the US National Weather Service.  For that purpose, it is run with a cubed-sphere grid of approximately \SI{13}{\km} horizontal spacing and 64 vertical levels.  For the results shown here, 
the grid spacing of the model is approximately \SI{140}{\km} resolution (C48) and there are 79 vertical levels.  
This coarser horizontal resolution greatly speeds up online testing of the emulated parameterization.

The training simulations are initialized from the GFS analysis from the first day of every month of 2016. 
The ZC inputs/outputs are saved every 5 hours in order to sample all times of day as the simulation progresses.
The simulations of Feb, June, and Sept are reserved for validation.
Overall the training set contains 1296 snapshots each consisting of $48^2\cdot 6 = \num{13824}$ atmospheric columns.

The net condensation $\Delta_g c$ at a given grid point $(x,y,z)$ in 3D space depends only on the thermodynamic inputs there, a property we refer to as grid-point locality. 
To exploit this locality, the \verb|gscond| subroutine is emulated with a single multi-layer perception which is applied to each grid-point separately.
The inputs and hyper-parameters are listed in Appendix \ref{sec:ml-inputs}.
Notably, the input set includes $T$, $q$, and pressure $p$.
In all the emulator takes in 14 scalar input channels and predicts a single scalar output: the net change $\Delta_g c$ in cloud condensate.
For online applications, we use the Fortran code instead of the emulator in the top 5 vertical levels since $\Delta_g c=0$ there for all training samples.
The results are similar when we directly enforce that $\Delta_g c = 0$.
Since $\Delta c$ depends exponentially on temperature we normalize it by the standard deviations computed within bins of temperature.
The corresponding change in specific humidity over the condensation step is $-\Delta_g c$, satisfying total water conservation.
The temperature change is $(L_v/c_p) \Delta_g c$ where $L_v$ is the latent heat of vaporization and $c_p$ is the specific heat of air at constant pressure. 
This is an approximation since some ZC phase changes occur between ice and vapor which releases additional latent heat.

For approximately 80\% of samples,  either the condensate change $\Delta_g c$ or the cloud after \verb|gscond| vanish. The regression model above makes small, but nonzero errors in these cases which accumulate in time and cause poor online performance.
To avoid this, we train a classifier to identify such points (see Appendix \ref{sec:classifier}) with the same inputs as the grid-point model described above.
This classifier is used online to either deactivate $\Delta_g c$, remove all the cloud, or return the grid-point model's output. 

The \verb+precpd+ subroutine is vertically non-local so we use a different architecture to predict $\Delta_p$ using the same input variables as the \verb+gscond+ emulator.
Because rain falls down rather than up, the predicted change at a given height of $T$, $q$, and $c$ depend only on data from overlying grid points.
We encode this causal assumption using an RNN which takes in a top-of-atmosphere to surface sequence of thermodynamic fields $\mathbb{R}^{79}\times\mathbb{R}^{14}$ and maps to $\Delta_p \in \mathbb{R}^{79}\times\mathbb{R}^3$.
The surface precipitation is also output as a trainable affine transformation of the final hidden RNN state (at the surface).
A two-layer vanilla RNN is used with ReLU activation, and 256 hidden nodes in each layer.
As a sensitivity test, we also train a two-layer column-dense model for this task with 256 hidden nodes in each layer.
Because heat and water conservation are not trivially enforced like they are for \verb|gscond| we predict the change of all three thermodynamic variables separately at the risk of violating appropriate conservation laws.  

All the models above are trained separately on the full training set with the Adam optimizer and a batch size of 512 columns. Mean-squared-error losses are used for regression models and cross-entropy is used for the \verb|gscond| classifier.  Hyper-parameters were not tuned exhaustively in this work apart from ensuring that the initial loss is $O(1)$ and stochastic gradient descent is stable. Both the precipitation and condensation regression models are trained by stochastic gradient descent over 25 epochs, with the Adam optimizer, a learning rate of \num{0.0001}, a batch size of \num{512} columns (\num{40448} grid points). The classifier is trained with an increased. learning rate of \num{0.001}.

All computations were performed on Google Cloud Platform. Training a model for 25 epochs takes 8 hours on a NVIDIA Tesla P4 GPU. On an 8 processor node, a 30 day FV3GFS simulation takes 5 hours to complete without ML and 10 hours with ML emulators.

\section{Results}\label{sec:results}


For each ML configuration, we run two simulations, one where the ML is active and another where the Fortran is active. 
The inactive scheme still saves data, so that we can compute skill metrics.
Metrics are computed over 30-day simulations initialized at the end of the June simulation from the training data.
Even though this overlaps in time with the July training data, it represents a unique testing dataset since it was initialized with the final state of the June training simulation rather than the July 1 GFS analysis.

We consider five metrics---the offline and online skill at predicting the $T$ and $c$ tendencies as well as the bias of the global mass of cloud water condensate per unit surface area (i.e. cloud water path) on July 15.
The Fortran-simulated cloud water path at this time was \SI{102}{\g\per\meter\squared}.
The tendency metrics measure the skill compared to the saved Fortran tendencies as given by a modified $R^2$ score $1-\sum (y-\tilde{y})^2/\sum y^2$, where $y$ is the truth and $\tilde{y}$ is the prediction.

We first examine the offline and online metrics for runs where the subroutines \verb|gscond| and \verb|precpd| are replaced by ML one-at-a-time (Table \ref{tab:table}).
Neural network architecture matters when replacing \verb|precpd|.
Using the RNN instead of a dense model improves the offline skill score for condensate tendency significantly, but the improvement in online skill is more dramatic:
the cloud bias is 20x worse with the dense model.
This large increase in cloud pushes the online states far from the envelope of the training data so the offline skill scores degrade.
This is a typical failure mode we have witnessed throughout this project.
Typically, the ML-predicted tendencies are fairly insensitive to state drift, but the Fortran physics produces very large tendencies as they attempt to correct small physical inconsistencies introduced by the ML such as super-saturated air or negative cloud concentrations.
The RNN avoids these deficiencies.

Using the classifier helps when replacing the just \verb|gscond| scheme (Table \ref{tab:table}, but the the impact is far more dramatic in an online simulation where \emph{both} \verb|precpd| and \verb|gscond| are replaced by ML emulators (Table \ref{tab:combined}).
In this run, \verb|precpd| is replaced by the RNN and \verb|gscond| is replaced with a dense-local model both with and without classifier-based masking.
Classifier-based masking is essential for good online skill.

\begin{figure}
    \centering
    \caption{(a) Zonal average of cloud water mixing ratio (mg/kg) from the truth and (b) bias of the emulation run. The average is computed from July 20 to July 31, inclusive. \label{fig:1}}
    \includegraphics{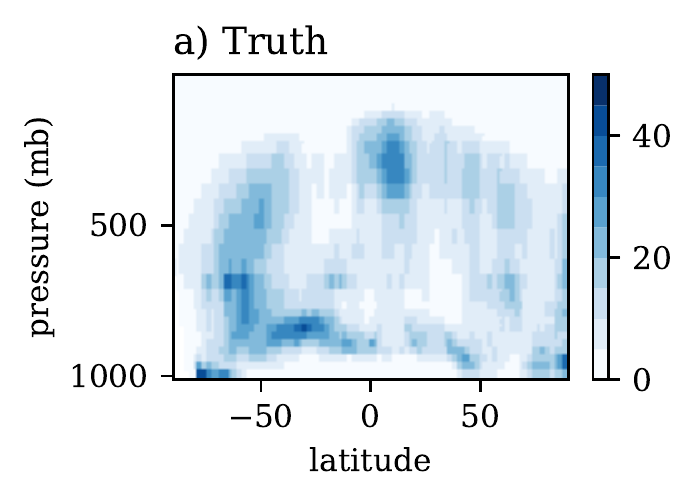}\includegraphics{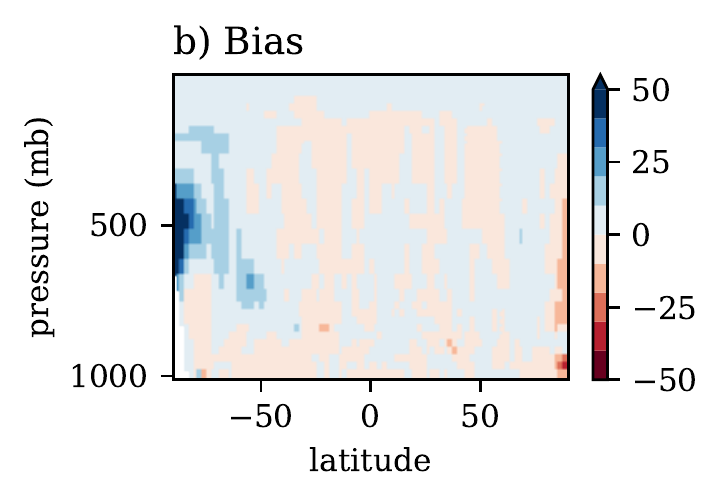}
\end{figure}

The combined ML simulation performs well by the skill metrics we have considered thus far, but it has excessive cloud over the Antarctic Plateau (Figure \ref{fig:1}).
The globally trained emulator also has poor offline skill there, so we tried using a separately trained condensation model for the high antarctic plateau.
This improved the offline skill, but not the online cloud biases
The interested reader can refer to Appendix \ref{sec:online} for more diagnostics of the online performance.

\begin{table}[]
\centering
        \caption{Table of skill scores for the \texttt{precpd} or \texttt{gscond} emulators. The online skill scores are computed in an online run where only the indicated subroutine is emulated.}
    \label{tab:table}
\begin{tabular}{@{}lllll@{}}
\toprule
                                 & \multicolumn{2}{c}{$\alpha = \texttt{precpd}$}  & \multicolumn{2}{c}{$\alpha = $\texttt{gscond}$$}    \\ 
                                 & Dense              & RNN       & No classifier     & Classifier \\
                                 \midrule
Offline $\Delta_{\alpha} c$ Skill              & 94\%               & 98\%      & 99\%              & 99.2\%     \\
Offline $\Delta_{\alpha} T$ Skill        & 92\%               & 98\%      & 98.50\%           & 98.7\%     \\
Online $\Delta_{\alpha} c$ Skill               & 14\%               & 95\%      & 92.60\%           & 93.9\%     \\
Online $\Delta_{\alpha} T$ Skill         & 68\%               & 98\%      & 92.50\%           & 94.1\%     \\
July 15 cloud water bias (g m$^{-2}$) & 23.6           & 1.3 & 25.0          & 21.2   \\ \bottomrule
\end{tabular}

\end{table}

\begin{table}[]
    \centering
    \caption{Online performance metrics for a simulation with emulators active for both \texttt{gscond} and \texttt{precpd}. The classifier predicts when cloud after \texttt{gscond} vanishes (Zero-cloud) and also when $\Delta_g = 0$. The columns show the impact of actually enforcing the predicted classes.}
    \begin{tabular}{@{}llll@{}}
    \toprule
         Metric & No Classifier & Enforce zero-cloud & Also enforce $\Delta_g = 0$ \\
\midrule
\verb+gscond+ skill ($\Delta_g c$) & -81\%         & -53\%                 & 94\%                      \\
\verb+precpd+ skill ($\Delta_p c$) & 36\%          & 36\%                  & 77\%                      \\
July 15 cloud water bias (g m$^{-2}$)  & 64.2      & 34.2              & 7.6                \\
\bottomrule
\end{tabular}
    \label{tab:combined}
\end{table}
The combined \verb|gscond|-\verb|precpd| model presented above required months of iterative development so we hope a brief description of our development trajectory will illuminate sensitivities we have not rigorously proven.
We began by attempting to emulate the combined action of \verb|gscond| and \verb|precpd| with a single dense neural network and then with RNNs.
The RNN was far better offline but worse online.
We then found that the \verb|gscond| subroutine used additional inputs: the humidity, surface pressure, and temperature after the last call to \verb|gscond| (Appendix \ref{sec:dataflow}).
Accounting for these improved the online performance of the RNN.
We temperature-scaled the condensation rate to improve offline skill in cold regions of Antarctica and the upper troposphere.
The largest improvements in online skill came from developing separate models for condensation and precipitation.
These schemes represent different physical processes and thus have different input-output structures and scales.

\section{Conclusions}\label{sec:conclusions}

Our results demonstrate that feasibility of emulating cloud processes of a climate model with a machine learning model.
Large-scale condensation is perhaps the fastest physical process in a climate model and we have developed the first emulator we know of it.
Our final model respected conservation laws, enforced the correct notion of grid-point locality, and employed classifiers to represent the mixed discrete-continuous structure of the predictand.
We also applied RNNs for the precipitation process in a novel way to enforce the fact that rain and snow fall down rather than up, leading to large improvements in offline and especially online skill.
This architecture could be useful for emulating other microphysical parameterizations, or more broadly parameterizations of other atmospheric processes that have a directional structure (e.g. orographic wave drag, entraining plumes).

Despite these successes, the detailed care required and the remaining issues over Antarctica temper our expectations for machine learning parameterization of processes with strong time-scale separation.
This manifests as a fat-tailed tendency targets which probably require high capacity models and vast training sets to learn.
For emulators, we can generate infinite training data but cannot use very high capacity models since emulators should be faster than the Fortran code they are emulating.
Machine learning parameterizations trained from observations or fine-resolution simulations face the further challenge of finite data and noise.

\begin{ack}
We thank Vulcan, Inc. and the Allen Institute for Artificial Intelligence for supporting this work. We acknowledge NOAA-EMC, NOAA-GFDL and the UFS Community for publicly hosting source code for the FV3GFS model and NOAA-EMC for providing the necessary forcing data to run FV3GFS.

The codes needed to produce the data and train models are available at \url{https://github.com/ai2cm/fv3net/tree/paper/microphysics/v1/projects/microphysics} with an MIT license.

A subset of the training data, some trained ML models, and the reduced data for the figures and tables is available on Zenodo at \url{https://doi.org/10.5281/zenodo.7109065} with a Creative Commons License.

\end{ack}

\section*{Broader Impacts}\label{sec:impacts}

Climate change and extreme weather impact us all.
This work helps illuminate a path for machine learning to improve weather and climate forecasts through ML emulation of physical parameterizations, which can in principle speed up the model and aid in the assimilation of new observations into the model.
Such forecasts help underpin strategies to adapt to and mitigate future climate change and extreme weather.
They also provide key scientific evidence for climate change and build the case for climate action.
While we cannot envision in immediate negative impacts of this work, our method could be applied to simulate physical systems we have not considered.  

\bibliography{main}
\bibliographystyle{plainnat}
\clearpage
\section*{Checklist}

\begin{enumerate}
\item For all authors...
\begin{enumerate}
  \item Do the main claims made in the abstract and introduction accurately reflect the paper's contributions and scope?
    \answerYes{}
  \item Did you describe the limitations of your work?
    \answerYes{}{See the last paragraphs of Sections \ref{sec:results} and \ref{sec:conclusions}}
  \item Did you discuss any potential negative societal impacts of your work?
    \answerYes{See ``Broader Impacts''.}
  \item Have you read the ethics review guidelines and ensured that your paper conforms to them?
    \answerYes{}
\end{enumerate}

\item If you are including theoretical results...
\begin{enumerate}
  \item Did you state the full set of assumptions of all theoretical results?
    \answerNA{}
	\item Did you include complete proofs of all theoretical results?
    \answerNA{}
\end{enumerate}

\item If you ran experiments...
\begin{enumerate}
  \item Did you include the code, data, and instructions needed to reproduce the main experimental results (either in the supplemental material or as a URL)?
    \answerYes{See the acknowledgements. Currently hidden to preserve anonymity.}
  \item Did you specify all the training details (e.g., data splits, hyperparameters, how they were chosen)?
    \answerYes{}
	\item Did you report error bars (e.g., with respect to the random seed after running experiments multiple times)?
    \answerNo{This is preliminary work and the sensitivies we discuss have large magnitudes.}
	\item Did you include the total amount of compute and the type of resources used (e.g., type of GPUs, internal cluster, or cloud provider)?
    \answerYes{See last paragraph of Section \ref{sec:methods}}
\end{enumerate}

\item If you are using existing assets (e.g., code, data, models) or curating/releasing new assets...
\begin{enumerate}
  \item If your work uses existing assets, did you cite the creators?
    \answerYes{We cited FV3GFS and Zhao Carr in Section \ref{sec:methods}.}
  \item Did you mention the license of the assets?
    \answerNA{}
  \item Did you include any new assets either in the supplemental material or as a URL?
    \answerYes{See acknowledgements.}
  \item Did you discuss whether and how consent was obtained from people whose data you're using/curating?
    \answerYes{See acknowledgements.}
  \item Did you discuss whether the data you are using/curating contains personally identifiable information or offensive content?
    \answerNA{}
\end{enumerate}

\item If you used crowdsourcing or conducted research with human subjects...
\begin{enumerate}
  \item Did you include the full text of instructions given to participants and screenshots, if applicable?
    \answerNA{}
  \item Did you describe any potential participant risks, with links to Institutional Review Board (IRB) approvals, if applicable?
    \answerNA{}
  \item Did you include the estimated hourly wage paid to participants and the total amount spent on participant compensation?
    \answerNA{}
\end{enumerate}

\end{enumerate}

\appendix

\clearpage
\section{The Zhao-Carr Microphysics}
\label{sec:zc}
This scheme handles both phase changes---condensation and evaporation---and precipitation processes. The former is typically 10x larger in magnitude. The prognostic variables used by the scheme are the temperature $T$, specific humidity $q$, and a combined cloud water/ice mixing ratio $c$.

The \verb|gscond| scheme handles evaporation of cloud and condensation.
Evaporation of cloud is given by $ E_c = \frac{1}{\Delta t}\max[\min[q_s(f_0 - f), c], 0]$.
$f$ is relative humidity. $f_0$ is  a critical relative humidity threshold which \citep{Zhao1997-hz} describe as ``was empirically set to 0.75 over land and 0.90 over ocean''.
$q_s$ is the saturation specific humidity.

Condensation $C_g$ on the other hand is given by a more complex formula involving a relative humidity tendency. See Eq. (8) of \citet{Zhao1997-hz}. 
Both formulas depend only on the thermodynamic state of a single $(x,y,z)$ location, but there is some non-local dependence on the assumed phase of the cloud and the corresponding latent heating rate.

The \verb|precpd| scheme handles the conversion of cloud into rain/snow and the evaporation of the latter as it falls through the atmosphere. Broadly speaking, it can be written as the following
\begin{align*}
E_{rr} = E_r(T, f, P_r)\\
E_{rs} = E_r(T, f, P_s)\\
P = P(T, f, c, P_r, P_s)\\
P_{sm} = P_{sm}(T, f, c, P_r, P_s)\\
P_r = \int_{p_t}^{p} (P - E_{rr}) dp/g \\
P_s = \int_{p_t}^{p} (P_{sm} - E_{rs}) dp/g.
\end{align*}
Most of the formulas are proportional to rainfall $P_r$ and snowfall $P_r$ rates at a given level, though are some rate constants that depend exponentially on temperature. $p_t$ is the pressure at the top of the atmosphere.
\clearpage
\section{Implementation of the Zhao-carr Microphysics in FV3GFS\label{sec:dataflow}}

\begin{figure}[h]
\begin{tikzpicture}[auto]

\tikzstyle{decision} = [diamond, draw, fill=blue!20, 
    text width=4.5em, text badly centered, node distance=3cm, inner sep=0pt]
\tikzstyle{block} = [rectangle, draw, fill=blue!20, 
    text width=10em, text centered, rounded corners, minimum height=4em]
\tikzstyle{line} = [draw, -latex']
\tikzstyle{cloud} = [draw, ellipse,fill=red!20, node distance=3cm,
    minimum height=2em]

\tikzstyle{var} = [align=right, text width=8em]

\node [var](tg) {$T$ after last call to gscond};
\node [var, below of = tg] (qg) {$c$ after last call to  gscond};
\node [var, below of = qg] (psg) {pressure after last call to gscond};
\node [var, below of = psg] (t) {$T$};
\node [var, below of = t] (qv) {$q$};
\node [var, below of = qv] (qc) {$c$};
\node [var, below of = qc] (ps) {pressure};

\draw  [->] (tg)  --  +(2.5, 0);
\draw  [->](qg)  --  +(2.5, 0);
\draw  [->](psg)  --  +(2.5, 0);

\draw [->] (t)  --  +(10, 0);
\draw [->] (qv)  --  +(10, 0);
\draw [->] (ps)  --  +(10, 0);
\draw [->] (qc)  --  +(10, 0);

\draw [->] (4.0, -3)  -- (4.0, 0) -- (10, 0);
\draw [->] (4.5, -4)  -- (4.5, -1) -- (10, -1);
\draw [->] (5, -6)  -- (5, -2) -- (10, -2);

\tikzstyle{sub}=[text width=8em, align=center];
\node [sub] (gscond) at (+3, 1) {Grid scale condensation \verb|gscond|};
\node [sub] (precpd) at (+6, 1) {Precipitation \verb|precpd|};
\node [sub] (precpd) at (+9, 1) {Rest of Model};

\draw[fill=blue] (2.5, 0.5) rectangle +(1, -6) ;
\draw[fill=blue] (5.5, -2.5) rectangle +(1, -3);
\draw[fill=gray] (8.5, -2.5) rectangle +(1, -4);

\draw[->] (ps.east) -| (3, -5.5);
\draw[->] (ps.east) -| (6, -5.5);

\end{tikzpicture}
\caption{Information flow of the {\color{blue} Zhao-carr microphysics} within FV3GFS. Inputs (outputs) of a given scheme are represented as inward (outward) arrows. The ``after last call to gscond'' inputs are used to compute a relative humidity tendency that encompasses the rest of the model and \texttt{prepcd}. This approach to computing the tendency effectively adds three new state variables to the model.}
\label{fig:schematic}
\end{figure}
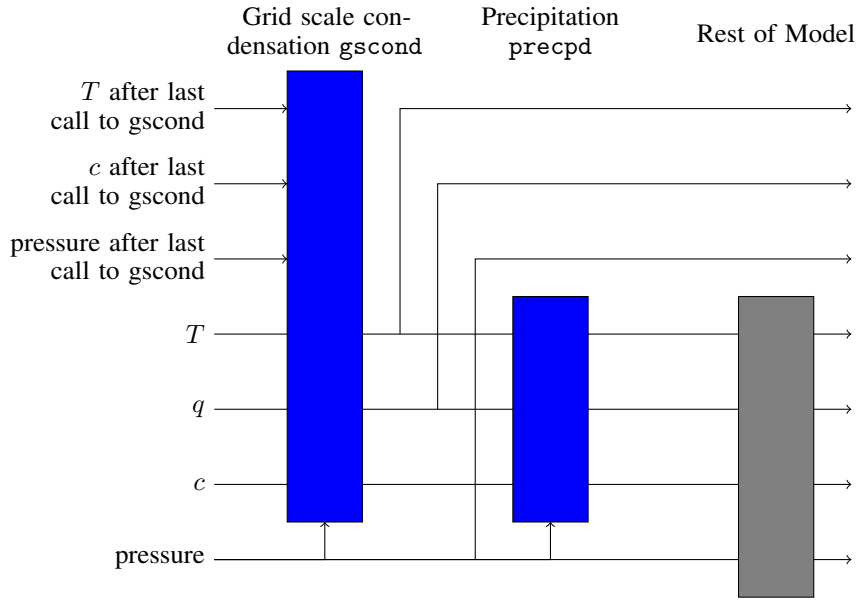
\clearpage
\section{Architectures and Hyper-parameters}
\label{sec:ml-methods}

\subsection{Input data and normalization}
\label{sec:ml-inputs}
The condensation emulator has 12 3d input variables: air pressure, air temperature, air temperature from the last time gscond was called (after-last gscond), cloud water mixing ratio, log of cloud, log of humidity, log of humidity after-last-gscond, pressure thickness of each atmospheric layer, specific humidity, specific humidity after-last-gscond. It has the following 2 2D inputs: surface air pressure and surface air pressure after-last-gscond. The Fortran implementation uses the after-last-gscond inputs to infer a relative humidity tendency.
Before being passed to the ML models the 2D inputs are replicated in the vertical direction, and all the variables are stacked along a channel dimension. 
The final ML input $x\in\mathbb{R}^{79}\times\mathbb{R}^{14}$. This input vector is then centered by removing the per-level and per-channel mean $\mu\in\mathbb{R}^{79}\times\mathbb{R}^{14}$ computed over a subset of samples. 
Then it is normalized by the standard deviation $\sigma\in\mathbb{R}^{14}$ computed over all vertical levels (79) and samples.

\subsection{Condensation}

\subsubsection{Regression Model}
The output of the condensation model is the change in cloud water $\Delta_g c$ over a single step of the \verb|gscond| subroutine. The condensation rate depends exponentially on temperature due to the Clausius-Clapeyron condition so we use a temperature-dependent normalization for the target.
We divide the range of temperatures into 50 equally spaced bins and compute the mean $\mu(T)$ and standard deviation $\sigma(T)$ of the target within each bin from a subset of the training data.
In initial tests, this improved the schemes performance in cold regions like the upper troposphere and Antarctica, though we have yet to show this with formal ablations of our finalized configuration.
Finally, the temperature scaled data are normalized once more with a per-level mean and all-level standard deviation.

The same neural network $f: \mathbb{R}^{12} \rightarrow \mathbb{R}$ is used for each level independently.
It is a multi-layer perceptron, with 2 hidden layers of 256 nodes each and ReLU activation.
It outputs the temperature-scaled target $\Delta_g c/\sigma(T)$. 
To ensure the model has skill both in temperature-scaled and the final output cloud $c_o = c_i + \Delta_g c$ we include both terms in the mean-squared error loss function given by
\[
\left|\frac{\Delta_g c-\mu(T)}{\sigma(T)} - f(x)\right|^2 + \lambda |\tilde{c_o} -c_o|^2.
\]
The predicted output cloud is $\tilde{c_o} = f(x) \sigma(T) + \mu(T)$.
The weight $\lambda = \num{50000} / Std[c_o]$ where $Std$ is the standard deviation computed from a subset of the training data.
This is chosen to so that both the predicted tendency and output cloud have $O(1)$ contributions to the loss.

\subsubsection{Classifier}
\label{sec:classifier}
For online application, we use a classifier to handle the mixed discrete-continuous nature of the target variable $\Delta_g c$.
It has the same architecture as the the regression model above, except with four target variables to identify the following classes:
\begin{itemize}
    \item $\Delta_g c = 0$ 
    \item cloud after \verb|gscond| vanishes but none of the above,
    \item $\Delta_g c>0$ but none of the above, and 
    \item $\Delta_g c<0$ but none of the above. 
\end{itemize}
In online/offline evaluations, if the classifier identifies the first two cases then we enforce the corresponding constraint on $\Delta_g c$.
Otherwise we use the regression model.
It is trained with a categorical cross entropy loss with the same hyperparameters as the regression models except for an increased learning rate of \num{0.001}.
After training the classifier is 98\% accurate over all classes and levels.
\subsection{Precipitation}
The precipitation emulator (\verb|precpd|) has the same input set and input normalization strategy as \verb|gscond|, the states at the beginning of the time-step. However, it predicts these 3D outputs: the change in air temperature, humidity, and cloud due to the precipitation subroutine. 
It also predicts surface precipitation rate which will moisten with the FV3GFS land-surface. 
All the outputs are scaled by the standard deviation $\sigma\in\mathbb{R}$ computed over all vertical levels. 
The loss function includes MSE terms for these direct predictands but also MSEs for the absolute values of cloud water, humidity, and air temperature at the end of \verb|precpd|. These MSEs are scaled empirically to have $O(1)$ contribution to the overall loss.

The precipitation emulator is vertically non-local unlike \verb|gscond|. To enforce the prior that rain falls down we use an recurrent neural network architecture (RNN). 
Let $x[i]\in\mathbb{R}^c$ be a vector with $c$ channels from level $i\in[0,79)$, where $i=0$ is the top of atmosphere and $i=78$ is the surface.
Similarly let $h[i]$ be a hidden state vector such that $h[i]=0$.
Then a single layer of the RNN is given by
$h[i+1] = (W_h h[i] + W_x x[i] + b)^+$, where $(\cdot)^+$ is the ReLU activation function.
We stack two such layers, with a hidden states of 256 channels.
Affine transformations map the final hidden state to the output fields  without mixing vertical levels like this $y[i]=Ah[i] + b$.
This architecture therefore ensures that $\nabla_{x[j]} y[i] = 0$ if $j > i$.

\clearpage
\section{Other aspects of online performance}
\label{sec:online}
\begin{figure}[h]
    \centering
    \includegraphics{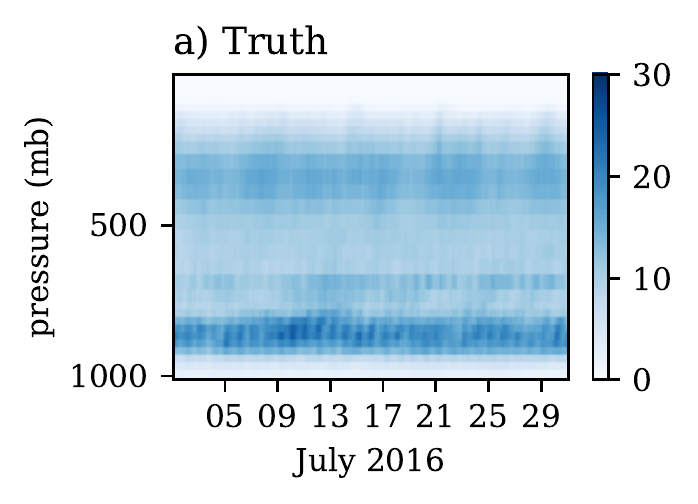}\includegraphics{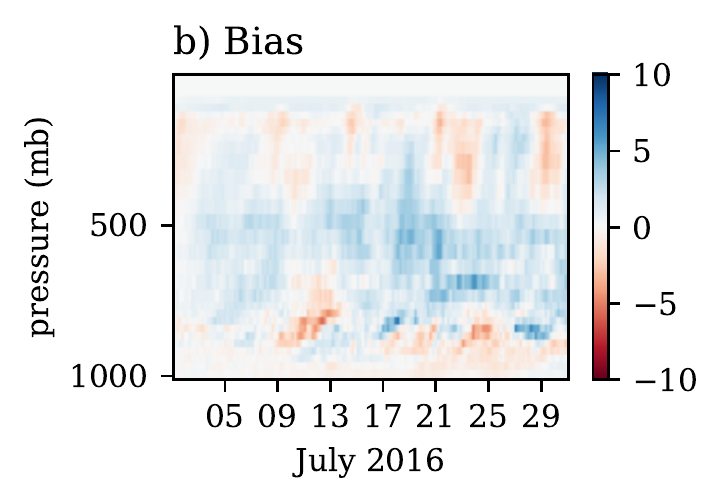}
    \caption{The global average cloud water mixing ratio (mg kg$^{-1}$) as a function of time and pressure. It shows  (a) the truth and (b) bias of the emulated run.}
    \label{fig:hovmoller}
\end{figure}

\end{document}